\documentclass[twocolumn,preprintnumbers,amsmath,amssymb]{revtex4}

\usepackage{epsfig}
\usepackage{graphicx}
\usepackage{dcolumn}
\usepackage{bm}

\usepackage[paperwidth=225mm,paperheight=295mm,centering,hmargin=1.8cm,vmargin=1.8cm]{geometry}

\usepackage{color}

\begin{document}
\title{SMILE Microscopy : fast and single-plane based super-resolution volume imaging
}

\author {Partha Pratim Mondal }
\email[Corresponding author: partha@iap.iisc.ernet.in]{}
\affiliation{%
Nanobioimaging Laboratory, Instrumentation and Applied Physics, Indian Institute of Science, Bangalore, INDIA \\
}%

\date{\today}
\begin{abstract}
Fast 3D super-resolution imaging is essential for decoding rapidly occurring biological processes. Encoding single molecules to their respective planes enable simultaneous multi-plane super-resolution volume imaging. This saves the data-acquisition time and as a consequence reduce radiation-dose that lead to photobleaching and other undesirable photochemical reactions. Detection and subsequent identification of the locus of individual molecule (both on the focal plane and off-focal planes) holds the key. Experimentally, this is achieved by accurate calibration of system PSF size and its natural spread in off-focal planes using sub-diffraction fluorescent beads. Subsequently the identification and sorting of single molecules that belong to different axial planes is carried out (by setting multiple cut-offs to respective PSFs). Simultaneous Multiplane Imaging based Localization Encoded (SMILE) microscopy technique eliminates the need for multiple z-plane scanning and thereby provides a truly simultaneous multiplane super-resolution imaging. 
\end{abstract}

\maketitle

\begin{figure*}
\includegraphics[height=3.5in,angle=0]{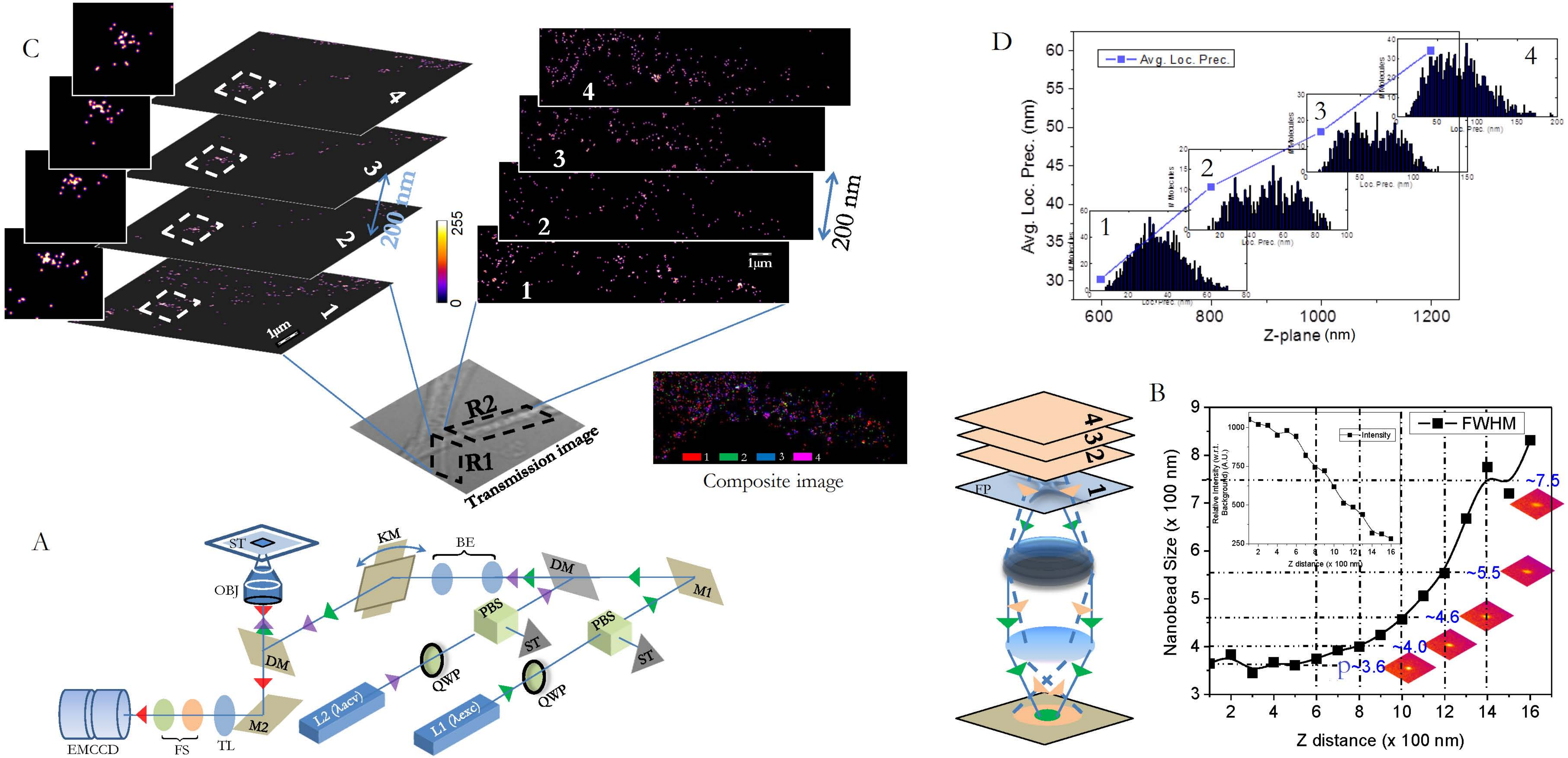}
\caption{\small{SMILE Microscopy: A. Optical diagram for SMILE microscopy; B. Mechanism behind SMILE microscopy along with the PSF size calibration curve obtained from near photobleachingless fluorescent nanobeads. The inset show the characteristic fall of intensity far from focus (at large z); C. Simultaneous imaging of multiple planes (of two different regions, R1 and R2) separated by $\Delta z = 200 ~nm$. The colored composite image (for region R2) that represent 3D positions (using plane numbers, 1-4) is also shown. The transmission image of the NIH-3T3 cell (Axon region) is also shown.  E. The characteristic increase in localization precision at large penetration depths indicating the role of scattering and absorption of emitted photons.  }}
\end{figure*}


Fast 3D super-resolution imaging is essential for visualizing rapidly occurring events in mammalian cells and microorganisms with sub-diffraction resolution \cite{parbook}\cite{coltharp}\cite{parSR}\cite{errington}. With the advent of single molecule based super-resolution imaging techniques, it is now possible to visualize and follow biophysical processes (protein dynamics and single molecule binding) within single mammalian cell, E. Coli and bacteria \cite{coltharp}\cite{errington}\cite{partha2012}. Prominent techniques that are capable of super-resolution include, photoactivated localization microscopy (PALM), fluorescence PALM (fPALM), stimulated emission depletion (STED), stochastic optical reconstruction microscopy (STORM), super-resolution optical fluctuation imaging (SOFI) and saturated structured illumination microscopy (SSIM) \cite{Superres}. Simultaneous Multiplane Imaging based Localization Encoded (SMILE) microscopy use PSF dimension to encode the location of single molecules in the respective z-planes. SMILE microscopy achieves this by determining the natural spread of PSF at greater depths, followed by single molecule sorting to enable simultaneous multiplane imaging. To construct volume image, existing techniques employ, light-sheet \cite{albyNM}\cite{betzig2016}, Bessel beam \cite{betzigBB}\cite{parthaSR} and cylindrical lens \cite{zhuang2008}. These techniques require sophisticated optical components (scanning mirror and Z-axis piezo stage) to render volume image \cite{parbook}. Moreover, some of these techniques are known to cause photobleaching and optical aberrations. A single-plane technique has the capability to eliminate these undesirable effects by simply reducing the radiation-dose and eliminating moving components. Here, we propose a single-plane super-resolution technique that utilizes the information (PSF size and number of emitted photons) from off-focal molecules. When integrated with the focal plane information it results in the reconstruction of 3D volume stack. We harness the information encoded in both diffraction-limited PSF and off-focal larger PSFs (Gaussian approximated). Due to diffraction nature of light, a point source (sub-diffraction nano-beads or fluorescent molecules) in focal plane appear as a diffraction-limited spot ($\Delta_{xy} \approx 1.22\lambda/2NA = 263 nm$) \cite{abbe}. However the spot size (PSF) broadens for point sources that are located in the off-focal planes (far from coverslip). This is due to diffraction and the spherical aberration (around the spherical beads) that results in bead size way larger than $\Delta_{xy}$ \cite{born1959}. Larger distances (typically, $>1.5\mu m$) gives rise to non-linear airy-disc like pattern that cannot be approximated by Gaussian function. Hence, we choose to work close to coverslip ($<1.5 \mu m$ from the coverslip). For generating simultaneous multiple planes from a single data set, we carried out a thorough calibration to experimentally obtain multiple cut-offs for system PSF size with desired inter-plane separation and plane thickness. These cut-offs decide the distance between adjacent Z-planes. This distinct size $W_z$ (as seen by the detector in the image plane) forms the basis for simultaneous multiplane imaging. Based on the calibration experiment, the following selection criterion (for selecting molecules $M_i$ in specific z-planes $Z_l$) is considered, $M_i \in Z_l ;~~l\Delta_{xy} < W_z < (l+1)\Delta_{xy} $ for $l$-planes and molecule $M_i$ with a size $W_z$. This criterion is applied on every detected single molecule (after localizing them) and a corresponding z-plane tag is assigned to the molecule. \\

The schematic diagram of SMILE microscopy is shown in Fig.1A. For calibration, sub-diffraction size nano-beads ($ < \Delta_{xy} $) embedded in Agarose transparent gel-matrix are imaged (Fig.1B). Nanobeads (point emitters) situated near the coverslip (focal plane of the detection objective) has diffraction-limited spot while those situated in other planes (slightly above the focal plane) have size $>\Delta_{xy}$. We study actin (labeled with a photoactivatable CAGE 552 dye) distribution in NIH 3T3 mouse fibroblast cells \cite{gudheti2013}. Experimentally recorded single molecule signatures can be visualized in the supplementary video. The study covers a distance of 1400 nm with a interplane separation of 200 nm starting from 600 nm above the coverslip. The region $<600 ~nm$ remain inaccessible due to negligible size (area $<3\times 3$ camera pixels) of the system PSF that cannot be reliably approximated by a 2D Gaussian function. Probing regions very close to coverslip ($<600 nm$) require high precision piezo z-stage that makes the system costly and complicated. We decided to keep the system simple yet reliable for 3D volume imaging. The separation is calibrated using sub-diffraction fluorescent nanobeads (Fig.1B) that determines the cut-offs for consecutive layers. Calibration shows a cutoff of $p\times p$ pixels (size of Gaussian approximated PSF) in the image plane (see, Fig1B). Sorting point sources (nano-beads) that belong to specific z-planes are then carried out based on their size using EINZEL Matlab scripts (kindly shared by, Prof. Samuel T. Hess, University of Maine, USA) \cite{Superres}\cite{gudheti2013}. Once z-dependent size of PSF is determined, the same becomes the reference for locating single molecules in the specimen planes. The transmission image of a part of single fibroblast cell along with the super-resolved images (regions, R1 and R2) and colored composite image for R2 are shown in Fig.1C. For both R1 and R2, 4 adjacent planes of super-resolved images are obtained simultaneously without the need for scanning individual planes. One can readily see change in the distribution of single molecules over 4 planes (spaced 200 nm apart) in the axon of fibroblast cells. Although some structures are restricted in the axial plane (i.e. only present in the first z slice adjacent to coverslip), others persist through planes at large depths. This is further evident from the colored composite image that represent the 3D position of single molecules in planes 1-4. Multiplane based SMILE imaging do suffer from scattering and photon absorption. This can be seen from the localization precision metric ($\Delta_{loc} = \Delta_{psf}/\sqrt{N}$, where $N$ is the number of photons detected for a single molecule) at large depths. Fig.1D shows the average localization precision for 4 different planes up to a depth of 1400 nm and the corresponding distribution of localization precision is also shown (see insets in Fig1D). The increase in average value of localization precision with increasing z-distance inside the specimen is quite evident. This reveals the fact that most of the photons emitted by the single molecule either gets absorbed by the adjoining nearby layers or gets scattered away thereby not reaching the detector. As a consequence of reduced detected photons, the average localization precision value increase at greater depths in the specimen. The increase is almost double for a depth (z-distance) of $\approx 600~nm$. This further explains the difficulty in realizing super-resolution at large penetration depths. The proposed SMILE microscopy is simple yet powerful technique that requires minimal optics and has the ability to reduce photobleaching thereby enabling single-plane 3D volume imaging (a step closer to temporal super-resolution). As a consequence of these advantages, one can now reliably image live cell in 3D and visualize rapidly occurring dynamical events in real-time. \\

\section*{Acknowledgment}
The author thank Prof. Samuel T. Hess (University of Maine, Orono, USA) for discussion and for sharing EINZEL analysis MATLAB codes that helped us develop new codes for SMILE microscopy. \\


\end{document}